\documentclass{PoS}

\usepackage{mathtools}
\usepackage{amsfonts}
\usepackage{amssymb}

\title{Gauge-independence of tunneling rates}

\ShortTitle{Gauge-independence of tunneling rates}

\author{Alexis D. Plascencia and \speaker{Carlos Tamarit} %
         \\
        Institute for Particle Physics Phenomenology\\
        Department of Physics, Durham University\\
        Durham DH1 3LE, United Kingdom\\
        E-mail: \email{a.d.plascencia-contreras@durham.ac.uk }, \email{carlos.tamarit@durham.ac.uk}}

\abstract{It is shown that tunneling rates can be defined in terms of a false-vacuum effective action whose reality and convexity properties differ from those of the corresponding groundstate functional. The tunneling 
rate is directly related to the false-vacuum effective action evaluated at an extremal ``quantum bounce''. The Nielsen identities of the false-vacuum functional ensure that the rate remains 
independent of the choice of gauge-fixing. Our results are nonperturbative and clarify issues related with convexity and radiative corrections.}

\FullConference{38th International Conference on High Energy Physics\\
                  3-10 August 2016\\
                 Chicago, USA}

\begin{document}
\section{Introduction}
\vskip-.1cm
In a quantum field theory with fields $\phi_i$, the effective action functional $\Gamma_{\rm eff}[\bar\phi_i]$, with $\bar \phi_i=\langle \phi_i\rangle_J$, encodes the dynamics of vacuum expectation values in the minimum energy state
and in the presence of external currents $J^i=-\delta\Gamma_{\rm eff}/\delta\bar\phi_i$. The zero momentum contribution to $\Gamma_{\rm eff}[\bar\phi_i]$ is the spacetime integral of the effective potential,
$V_{\rm eff}[\bar\phi_i]$.  The latter is a very relevant object, as it determines the  structure of Lorentz-symmetry-preserving vacua in the theory: for constant expectation values and in the absence of 
currents, $J_i=-\delta\Gamma_{\rm eff}/\delta\bar\phi_i=0$ implies $\delta V_{\rm eff}/\delta\bar\phi_i=0$. That is, the expectation values of fields in the vacuum minimize the effective potential.

Famous examples in which the effective potential has been used to learn about  new vacua are models with radiative symmetry breaking \cite{Coleman:1973jx}, or the Standard Model with its radiative Higgs instability  \cite{Hung:1979bg}.
However, the effective potential is plagued by some potentially problematic issues. First, in gauge theories it is known to depend on the choice of gauge-fixing \cite{Jackiw:1974cv}. Reassuringly, this gauge dependence disappears at the minima \cite{Nielsen:1975fs}, but when it comes to observables that are sensitive to the potential in between them, such as tunneling rates, the situation has remained unclear.
A second problem is that formal proofs show that the effective
potential is a concave function of the fields, meaning that its second derivatives  are always positive or zero \cite{Iliopoulos:1974ur}. This seems to be in conflict with the shape of the potential in perturbative calculations.
For example, the Higgs instability arises after the potential at large field values starts to decrease, becoming convex (with a negative second derivative). Finally, connected with the above, there are further
issues related again with tunneling rates. On the one hand, if vacua are determined from the effective potential, but the latter is concave, how can there be multiple vacua?  And finally, in models in which new
vacua arise radiatively --such as in the examples alluded to before-- how can one consistently calculate the tunneling rate? The traditional method \`a la Callan-Coleman \cite{Callan:1977pt}  relies on perturbing around
a bounce solution to the \emph{classical} Euclidean equations of motion. If the new vacua appear radiatively, there is no such solution. In the scientific literature it has been frequently assumed 
that the effective potential hast to play a role, but then the problems with convexity and gauge-dependence must be addressed. Alternatively, one can consider effective theories with gauge degrees of freedom integrated out \cite{Metaxas:1995ab}, but then the connection with the full effective potential disappears  and the issue of gauge-independence at all orders remains obscure.

As noted in \cite{Plascencia:2015pga}, whose results are summarized by these proceedings, all the previous problems can be clarified by realizing that there is a distinction between functionals of expectation values in the groundstate,
and analogue functionals of expectation values in a false vacuum. The false vacuum is not an eigenstate of the Hamiltonian, and as a consequence of this the functionals derived from its 
transition amplitude have different convexity and reality properties than their true vacuum analogues. In this way, the previous obstructions preventing the groundstate effective action from playing a role in the description of tunneling processes do not apply for the false-vacuum effective action, from which tunneling rates can be extracted in a way that guarantees gauge-independence. This is
explained in the following sections, first focusing on effective action functionals, and then moving on into their gauge-dependence and tunneling rates.

\section{Effective action functionals}
\vskip-.1cm
The groundstate effective action is built from the true-vacuum partition function $Z[J]$ in the presence of a source $J$. This functional gives the  probability amplitude of the transition from the groundstate onto itself.
One can consider a basis of time-independent eigenstates $|q\rangle$ of the field operators in the Heisenberg picture, satisfying $\hat\phi_i|q\rangle=q_i|q\rangle$. Using the spectral decomposition of the identity into
projection operators $| q\rangle\langle q |$, one can write $Z[J]$ as
\begin{equation}
\label{eq:W0p}
\begin{aligned}
 Z[J]\equiv\exp i W[J]=\lim_{T\rightarrow\infty}\langle 0| e^{-iHT}|0\rangle^J=\lim_{T\rightarrow\infty}\int [dq] [dq']\langle0|q\rangle^J\langle q|e^{-iHT}|q'\rangle^J\langle q'|0\rangle^J\\
 =\int [dq][ dq']\psi^J_0(q')\psi^{J\star}_0(q)\int_{q'}^q[d\phi]\exp{i\left[\tilde S_g[\phi;\xi]+J_j\phi^j\right]}\equiv\langle\exp[i J_j\phi^j]\rangle.
 \end{aligned}
\end{equation}
In the above expressions, $\tilde S_g[\phi;\xi]$ denotes  the action plus gauge-fixing terms, and $\xi$ is the gauge-fixing parameter. The contraction $J_j\phi^j$ represents a sum over different fields, with an integration over spacetime indices. $\psi^{J}_0(q)\equiv \langle q |0\rangle^J$ designates the wave function of the groundstate in field space. The vacuum transition amplitude $Z[J]$ can be written as $\lim_{T\rightarrow\infty}\exp(-iE_0 T)$, where $E_0$ is the energy of the groundstate, and its analytic continuation to imaginary time defines a real functional which formally looks like an integration with a positive real measure. Then H\"older's inequality can be used to prove that $W[J]$ must be a concave functional of $J$ \cite{Iliopoulos:1974ur}.

Usually, the integration over the vacuum wave function in \eqref{eq:W0p} is ignored, and the last path integral by itself, with some given boundary conditions $q=q'=q_0$,  is taken as a definition of the partition function. This is equivalent to considering a narrowly peaked wave function, $\psi^{J}_0(q)=\delta(q-q_0)$. However, in the presence of false vacua one expects the previous approximation to fail. The true groundstate should be a superposition of the perturbative groundstates associated with the false vacua, and so its wave function can  spread out in field space. Even if in the absence of currents there is a clearly preferred vacuum, thus giving a localized wave function,  nonzero currents $J_i$ can displace the energies of the vacua, so that for different values  of the $J_i$ the true-vacuum wave function will either localize in different regions of field space --when one vacuum dominates-- or spread out when several vacua compete. Hence a better approximation to the partition function can be obtained by assuming a multi-peaked structure of the wave-function with current-dependent coefficients, e.g. $\psi^{J}_0(q)=\sum_m{\cal C}_m\delta(q-q_0^{J,m})$. This implies that $Z[J]$ is given by a sum of multiple path integrals with different boundary conditions: 
\begin{align}
 Z[J]\approx{\sum_{m, n=1}^N }Z^{m,n}[J],\quad Z^{m,n}[J]= {{\cal N}[J]}_{mn} \int_{q^{J,m}_0}^{q^{J,n}_0}[d\phi]\exp{i\left[ \tilde S_g[\phi;\xi]+J_j\phi^j\right]}.
\end{align}
From the partition function $Z[J]$ one can construct the groundstate effective action as a Legendre transformation: $\Gamma[\bar\phi]=-i\log Z[J]-J_j\,\bar\phi^j$, with $J_i=-\delta\Gamma/\delta\phi^i\equiv-\Gamma_{,j}$. $\Gamma$ inherits the convexity properties of $W$, which results in a concave effective potential. Again, in the presence of multiple vacua $\Gamma$ will be determined by a sum of path integrals, which, assuming currents that enforce a groundstate with $\langle\phi\rangle=\bar\phi_\infty$ at infinite times, can be written as
\begin{align}
\label{eq:Gamma}
\exp i\Gamma[\bar\phi;\xi]=&{\sum_{m,n=1}^N}{{\cal N}[\Gamma_{,j}]}_{mn}\int_{q^{J,m}_0-\bar\phi_\infty}^{q^{J,n}_0-\bar\phi_\infty} [d\phi] \,\exp i\left[S_g[\bar\phi,\phi;\xi]-\Gamma_{,j}[\bar\phi;\xi]\phi^j\right],
\end{align}
with $S_g[\bar\phi,\phi;\xi]=\tilde S_g[\bar\phi+\phi;\xi]$. There are explicit examples in the literature in which, in the presence of multiple vacua, a single path integral fails to yield a concave effective potential, and yet summing over path integrals with different boundary conditions yields a concave one \cite{Fujimoto:1982tc}. In our formalism we interpret this need to sum over path integrals as a natural consequence of a multi-peaked wave function, rather than arising from sums over semiclassical paths; also, we stress the necessity of ``off-diagonal'' boundary conditions related to interference between the false groundstates.

Similarly to what was done for the groundstate, one may define a partition function $Z^T_F[J]=\exp i W^T_F[J]=\langle F|\exp(-iHT)|F\rangle$ for a false vacuum $|F\rangle$, and its associated effective action. There are important differences with the groundstate functionals. First, as $|F\rangle$ is not an eigenstate of the Hamiltonian, the transition amplitude $Z^T_F[J]$ is complex valued. One can define
the energy density $\epsilon$ of the state $|F\rangle$ by writing $\langle F|\exp(-iHT)|F\rangle=\exp(-i\epsilon VT)$. For an unstable, decaying state, $\epsilon$ has an imaginary part, which gives an exponentially decaying contribution to the false-vacuum amplitude related with the decay rate $\gamma$:
\begin{align}
\label{eq:rate}
 \gamma=-2\,{\rm Im}\epsilon=-\lim_{V,T\rightarrow\infty}\frac{2}{VT}\,{\rm Re}\,(\log Z^T_F[0]).
\end{align}
The imaginary part of the energy density prevents relating $Z^T_F[J]$ with a real functional by analytic continuation to imaginary time. Hence, $Z^T_F[J]$ cannot be written as
an average of a positive real functional, so that the arguments used to prove concavity of the groundstate functional $W[J]$ do not apply to $W_F[J]$. On the other hand, false vacua, not being eigenstates of the full theory, can be thought of as perturbative states of a theory expanded around some local energy minimum. Therefore they should have a localized wave-function in field space, which justifies a single path-integral approximation for $Z^T_F[J]$, unlike for $Z[J]$. As done before for the groundstate, one may define a false vacuum effective action
$\Gamma_F$ by means of
a Legendre transformation. As opposed to $\Gamma$, $\Gamma_F$ will be well approximated by a single path integral, and it will be complex. Furthermore, its effective potential will not be concave
and can exhibit multiple minima. Analogously to \eqref{eq:Gamma},  $\Gamma_F$ can be defined as
\begin{align}
\label{eq:GammaFLegendre}
 \Gamma^T_F[\bar\phi]=-i\log Z^T[J]- J_j\,\bar\phi^j,\quad J^i=-\frac{\delta\Gamma^T_F}{\delta\phi_i},
\end{align}
and is given implicitly by the following expression in the single path integral approximation:
\begin{align}
\exp i\Gamma^T_F[\bar\phi;\xi]=&\int_{0}^{0} [d\phi] \,\exp i\left[S_g[\bar\phi,\phi;\xi]-\Gamma^T_{F,j}[\bar\phi;\xi]\phi^j\right].
\end{align}

\section{Tunneling and gauge dependence}
\vskip-.1cm
As anticipated before, the effective action $\Gamma$ is known to depend on the choice of gauge parameter $\xi$. This gauge dependence is encoded by Nielsen identities \cite{Nielsen:1975fs}, which can be derived by means
of path-integral methods. The derivation can be applied to either the groundstate or the false-vacuum effective action, with the resulting identities having the form (either for $\Gamma$ or $\Gamma_F$)
\begin{align}
\label{eq:Nielsen}
 \xi\frac{\partial\Gamma}{\partial\xi}[\bar\phi;\xi]+\frac{\delta\Gamma}{\delta\phi_j}[\bar\phi;\xi] K^j[\bar\phi,\xi]=0,
\end{align}
where $K^j$ is a functional given by the quantum average of a gauge-transformation that depends on the details of the gauge fixing. Crucially, the dependence on the gauge parameter disappears
when the effective actions are evaluated on solutions to the quantum equations of motion $\delta\Gamma/\delta\phi_j=0$. This happens for example at the minima of effective potential, which are stationary points of the effective action. 
The false vacuum effective potential, which is not concave, can have several minima, and the false vacuum effective action can have stationary field trajectories that interpolate between vacua. 
It turns out that the tunneling rate can be related to $\Gamma_F$ evaluated at one of these extrema, so that Nielsen identities guarantee a result that does not depend on the choice of gauge fixing. Equation \eqref{eq:GammaFLegendre} implies that, in the absence of a current $J^i=-\delta\Gamma^T_F/\delta\phi_i=0$, then
$\Gamma^T_F[\varphi]=-i\log Z^T_F[0]$, where the condition of zero external current implies that $\varphi$ must be an extremal configuration. Then, using the definition \eqref{eq:rate} of the tunneling rate, it follows that 
\begin{equation}
\gamma=\frac{2}{VT}\,{\rm Im}\,\Gamma^T_F[\varphi_F(\xi);\xi], \quad \text{with} \quad \frac{\delta\Gamma^T_F}{\delta\phi_i}[\phi;\xi]|_{\phi=\varphi_F(\xi)}=0,
\end{equation}
where we emphasized that the extremal solution is itself gauge-dependent. Despite this, the Nielsen identities ensure gauge-independence of the tunneling rate.

So far we did not specify the solutions to the quantum equations of motion of $\Gamma_F$, or their boundary conditions. Going back to the path integral definition of $Z^T_F[0]$, it can be shown that the tunneling rate involves a sum over solutions which
tend to the false-vacuum configuration $\phi\equiv q_F$ at $t=\pm T/2$. These solutions can bounce back and forth from the false vacuum any integer number of times, and if the bounces are well separated in time, one can sum their contributions, which give the exponential of the single-bounce result \cite{Plascencia:2015pga}. After rotating to Euclidean time, the tunneling rate is then
\begin{align}
\label{eq:gammafinal}
\gamma=\frac{2}{VT^E}\,{\rm Im}\,\, e^{-\Gamma^{E}_F[{\varphi^{1,E}(\xi);}\xi]}
={2\cal J}\,{\rm Im}\,\, e^{-\Gamma'^{E}_F[{\varphi^{1,E}(\xi);}\xi]}.
\end{align}
$V$ and $T^E$ represent spatial volume and Euclidean time,  $\varphi^{1,E}(\xi)$ denotes the Euclidean single-bounce solution, and ${\cal J}$ is a Jacobian associated with space-time translations (zero modes) of the 
single bounce.  ${\Gamma'_F[\varphi^{1,E}(\xi)]}$ is the effective action evaluated at the bounce, without the integration over its zero modes. At lowest order in a derivative expansion, the Jacobian can be shown to be 
\begin{align}
\label{eq:Jexp}
 {\cal J}\sim\left[\frac{{\Gamma'}^{E}_F[\varphi^{1,E}(\xi);\xi]}{2\pi V''_{\rm eff}(q_F)}\right]^2.
\end{align}
 Equations \eqref{eq:gammafinal} and \eqref{eq:Jexp} make gauge-independence explicit, and provide a nonperturbative generalization of Callan and Coleman's semiclassical results \cite{Callan:1977pt}. In Callan and Coleman's formula, the tunneling rate involves the exponential of the  classical Euclidean action $S_E$ evaluated on a bounce solution, as well as a Jacobian proportional to $S^2_E$, and a fluctuation determinant. Our results \eqref{eq:gammafinal} and \eqref{eq:Jexp} show that the fluctuation determinant can be resummed, with the effect of replacing the classical action with the false-vacuum effective action. Moreover, our formalism shows that, rather than perturbing around
a classical bounce solution, one has to consider the full quantum bounce, as in the approach of \cite{Garbrecht:2015oea}. Finally, expressing the tunneling rate in terms of the false-vacuum effective
action and its solutions clarifies how to proceed when vacua or instabilities arise radiatively.

\section{Conclusion}

We have shown that tunneling rates can be expressed directly in terms of the imaginary part of a false-vacuum effective action evaluated at an extremal solution --a ``quantum bounce''-- with boundary conditions fixed by the former vacuum. The result, which generalizes Callan and Coleman's tunneling formula, is nonperturbative and ensures gauge-independence of the tunneling rate, which arises as a consequence of the Nielsen identities for the false-vacuum functional. 

The notion of  false-vacuum effective action, which generalizes the concept of ``localized'' effective action in \cite{Weinberg:1987vp}, clarifies some puzzles related with the known reality and convexity properties of the groundstate effective action. Such properties prevent the latter functional from playing a role in tunneling rates. The false-vacuum effective potential is not required to be concave, and can thus exhibit several minima. Consequently, the false-vacuum effective action can support bounce solutions interpolating between vacua, in contrast to the groundstate action, whose concave potential can only have a single minimum. Moreover, the false-vacuum effective action is complex, and its imaginary part --coming in part from convex regions 
in the effective potential, as can be argued from unitarity \cite{Weinberg:1987vp}-- encodes the tunneling rate, as follows from expression \eqref{eq:gammafinal}. 

We have argued that, in contrast to the groundstate effective action in the presence of several vacua, the false-vacuum effective action can be well approximated by a single path integral. This explains the apparent contradiction between the known convexity properties of the groundstate action and explicit computations returning convex effective potentials, as in the celebrated case of the Standard Model  instability. There is no contradiction;  rather, the calculations, obtained from a single path integral, correspond to a false-vacuum effective action associated with the perturbative vacuum, and their results are still meaningful.

Finally, our formula for the tunneling rate in terms of the false-vacuum effective action and its bounce solutions clarifies how to compute tunneling rates when the false vacuum arises
radiatively and, absent a classical bounce, Callan and Coleman's formula cannot be applied.


\begin{thebibliography}{99}

\bibitem{Coleman:1973jx}
  S.~R.~Coleman and E.~J.~Weinberg,
  Phys.\ Rev.\ D {\bf 7} (1973) 1888.
  doi:10.1103/PhysRevD.7.1888
  
\bibitem{Hung:1979bg}
  P.~Q.~Hung,
  FERMILAB-PUB-79-014-T-REV, FERMILAB-PUB-79-014-THY-REV.


\bibitem{Nielsen:1975fs}
  N.~K.~Nielsen,
  Nucl.\ Phys.\ B {\bf 101} (1975) 173.
  doi:10.1016/0550-3213(75)90301-6

\bibitem{Iliopoulos:1974ur}
  J.~Iliopoulos, C.~Itzykson and A.~Martin,
  Rev.\ Mod.\ Phys.\  {\bf 47} (1975) 165.
  doi:10.1103/RevModPhys.47.165

\bibitem{Jackiw:1974cv}
  R.~Jackiw,
  Phys.\ Rev.\ D {\bf 9} (1974) 1686.
  doi:10.1103/PhysRevD.9.1686

  
\bibitem{Callan:1977pt}
  C.~G.~Callan, Jr. and S.~R.~Coleman,
  Phys.\ Rev.\ D {\bf 16} (1977) 1762.
  doi:10.1103/PhysRevD.16.1762
  
\bibitem{Metaxas:1995ab}
  D.~Metaxas and E.~J.~Weinberg,
  Phys.\ Rev.\ D {\bf 53} (1996) 836
  doi:10.1103/PhysRevD.53.836
  [hep-ph/9507381].
  
  
\bibitem{Plascencia:2015pga}
  A.~D.~Plascencia and C.~Tamarit,
  JHEP {\bf 1610} (2016) 099
  doi:10.1007/JHEP10(2016)099
  [arXiv:1510.07613 [hep-ph]].
  
\bibitem{Fujimoto:1982tc}
  Y.~Fujimoto, L.~O'Raifeartaigh and G.~Parravicini,
  Nucl.\ Phys.\ B {\bf 212} (1983) 268.
  doi:10.1016/0550-3213(83)90305-X

\bibitem{Weinberg:1987vp}
  E.~J.~Weinberg and A.~q.~Wu,
  Phys.\ Rev.\ D {\bf 36} (1987) 2474.
  doi:10.1103/PhysRevD.36.2474
  
\bibitem{Garbrecht:2015oea}
  B.~Garbrecht and P.~Millington,
  Phys.\ Rev.\ D {\bf 91} (2015) 105021
  doi:10.1103/PhysRevD.91.105021
  [arXiv:1501.07466 [hep-th]].
\end{thebibliography}
\end{document}